\documentclass[11pt,a4paper]{article}
\usepackage{jstyle, enumitem}
\usepackage{hyperref}
\usepackage{slashed,framed}
\usepackage{empheq}
\usetikzlibrary{calc,decorations.markings}
\DeclareMathOperator{\csch}{csch}

\author[a]{Jin-Beom BAE}
\author[b,c]{\quad Euihun JOUNG}
\author[d]{\quad Shailesh LAL}

\affiliation[a]{Korea Institute for Advanced Study, 85 Hoegiro, Dongdaemun-Gu, Seoul 02455, Korea}
\affiliation[b]{School of Physics and Astronomy, Seoul National University, Seoul 151-747, Korea}
\affiliation[c]{Gauge, Gravity \& Strings, Center for Theoretical Physics of the Universe, Institute for Basic Sciences, Daejeon 34047, Korea }
\affiliation[d]{LPTHE -- UMR 7589, UPMC Paris 06, Sorbonne Universit{\'e}s,  Paris 75005, France}

\emailAdd{jinbeom@kias.re.kr}
\emailAdd{euihun.joung@snu.ac.kr}
\emailAdd{shailesh@lpthe.jussieu.fr}

\begin{document}

\title{\centering A Note on Vectorial AdS$_5$/CFT$_4$
Duality \\ for Spin-$j$ Boundary Theory}

\abstract{The vectorial holographic correspondences between higher-spin theories in AdS$_5$ and  free vector models on the boundary are extended to the cases
where the latter is described by free massless spin-$j$ field.
The dual higher-spin theory in the bulk does not include gravity and
can only be defined on rigid AdS$_5$ background with $S^4$ boundary.
We discuss various properties of these rather special higher-spin theories 
and calculate their one-loop free energies.
We show that the result is proportional to the same quantity for spin-$j$ doubleton treated as if it is a AdS$_5$ field.
Finally, we consider even more special case
where the boundary theory itself is given by
an infinite tower of massless higher-spin fields.
}


\maketitle

\section{Introduction}

The holographic dualities where the boundary theories carry vector representations
of $U(N)$ or $O(N)$ have been actively studied in recent years.
These vector models are conjectured to be dual to
Vasiliev's higher spin gravity \cite{Vasiliev:1990en,Prokushkin:1998bq} 
in the bulk. The proposal was originally made for AdS dimensions four or higher \cite{Sezgin:2001zs,Sezgin:2001yf,Sezgin:2001ij,Sezgin:2002rt,Klebanov:2002ja}
and  then extended to AdS$_3$/CFT$_2$ in \cite{Gaberdiel:2010pz}. 
We refer the reader to the reviews \cite{Giombi:2012ms,Giombi:2016ejx} for further details and progresses in this area.
The vectorial dualities has simple classical bulk spectra which can
be exactly identified.
This  special feature makes it possible to compute the one-loop partition function
of the bulk theory by
simply collecting the quantities for individual fields  
\cite{Camporesi:1990wm,Camporesi:1994ga,Camporesi:1995fb,Gopakumar:2011qs}
and led to possible insights into 
the existence of multiparticle symmetries of the theory \cite{Gaberdiel:2010ar,Gupta:2012he}.
In general dimensions, the one-loop partition function has been computed 
for the type A Vasiliev theory dual to free scalar on the boundary in \cite{Giombi:2013fka, Giombi:2014iua, Giombi:2014yra}. 
In addition to the free scalar/type A duality, 
two other dualities can be considered in  AdS$_5$/CFT$_4$
whose boundary theories are described by free fermion and Maxwell fields, respectively.
The corresponding five-dimensional bulk theories are refered to as type B and C  higher-spin theories. 
The spectific bulk description is known only for type-A case:
it is described by the any-$d$ Vasiliev equation \cite{Vasiliev:2003ev}.
About the construction of type-B higher-spin theory, see \cite{Vasiliev:2004cm}. Even though we do not know exact formulations of type B and C  theories, their perturbative field contents can be identified
from the correspondence, and the one-loop partition function 
was computed in \cite{Beccaria:2014zma}.
See \cite{Beccaria:2014xda,Beccaria:2014qea,Beccaria:2015vaa,Beccaria:2016tqy,Pang:2016ofv,Gunaydin:2016amv,Giombi:2016pvg} 
for other discussions and more recent developments on the vectorial dualities,
and \cite{Bae:2016rgm,Bae:2016hfy} for
the extension of the one-loop partition function computation
to free scalar and Yang-Mills adjoint models.

Motivated by these developments, in this paper, we consider a straightforward but in a sense rather exotic generalization of type A, B and C dualities in AdS$_5$/CFT$_4$\,. 
The generalization is based on the fact that the short representations of four-dimensional conformal symmetry $so(2,4)$ 
are the spin-$j$ massless particles  where $j$ is any half-integer number. 
Clearly the cases of $j=0,\frac12$ and 1 give the type A, B and C models,
whereas the other cases with $j\ge\frac 32$ have not been studied much
and there are good reasons for that.
See the appendix C of \cite{Gunaydin:2016amv} for 
previous discussions on the point in the literature.
Massless fields with spin $j\ge \frac32$ can be defined in a conformally flat and 
Einstein background \cite{Christensen:1978md,Aragone:1979hx}. 
Hence, one can consider $S^4$ as a consistent background, but another typical background
$S^1\times S^3$ is not compatible.
Even though the spin-$j$ theory on $S^4$ admits  global conformal symmetry
--- its Hilbert space carry $so(2,4)$ unitary representation --- 
it does not have a local energy momentum tensor.
Moreover, the symmetry cannot be realized at the level of gauge potential but only curvature \cite{Barnich2015}, hence for instance, the conformal symmetry 
cannot be realized in the standard Fronsdal formulation at least locally.
Despite of these aspects, as we shall show, single-trace operators constructed from Fronsdal fields do carry proper representations of conformal algebra. 
This encourages us to consider the holographic dualities based on the vector models of
massless spin-$j$ field,
whose bulk dual theory we shall refer to as  `type-$j$' higher-spin theory in analogy to type A, B and C. 
The type-$j$ theories have another exotic property that
they do not involve any massless spin-two field.
Hence, these theories are non-gravitational higher-spin gauge theories in rigid AdS$_5$ background.
This generalized version of AdS/CFT duality have been already considered in
different contexts \cite{Aharony:2015zea,Paulos:2016fap}.
We also analyze an even more exotic but interesting model where the boundary theory involves all massless higher-spin fields. 
Effectively the boundary theory is the free limit of four-dimensional Vasiliev theory defined now on the $S^4$. 

The rest of the present paper is organized as follows. 
In Section \ref{sec:typej}, we briefly review the spin-$j$ doubleton representations, their character formulae and the single-trace operators. 
In Section \ref{sec:classicalHS}, we move to the AdS$_5$ side and discuss classical aspects of 
 the bulk duals of the spin-$j$ vector models, including their field content, cubic interactions and higher-spin symmetries.
In Section \ref{sec:oneloopHS}, 
we calculate the one-loop vacuum energy of type-$j$ higher-spin theory
and the bulk theory dual to all massless integral spins.
We also comment on the Casimir energy.
The final section \ref{conclusion} 
contains discussions.

\section{Vector Model of Spin-$j$ CFT$_4$}
\label{sec:typej}

\subsection{Massless Spin-$j$ Theory}

In four dimensions, free massless fields of any spin carry representations of 
conformal symmetry $so(2,4)$ and they correspond to the 
UIR $\cD(j+1,(j,\pm j))$\,,
where $\cD(\D,(\ell_1,\ell_2))$ is the UIR 
having lowest energy $\D$ and its eigenvector(or tensor) 
transforms as  $(\ell_1,\ell_2)$ Young diagram representation of $so(4)$\,.
In the notation of $su(2,2)$
they are denoted by $\cD(j+1,[j,0])$ or $\cD(j+1,[0,j])$
where $[j_+,j_-]$ is the $(2\,j_++1)\times (2\,j_-+1)$ dimensional representation
 of $su(2)\oplus su(2)\subset su(2,2)$\,.
These representations are often referred as to spin-$j$ \emph{singleton} or \emph{doubleton}.
The sign of $\ell_2$ or asymmetry in $j_+$ and $j_-$ reflects that
representations are chiral or anti-chiral
whose field theoretical realization would require complexification of fields.
Here, we consider parity-invariant combination,
\be
	\cS_j:=\cS_{[j,0]}\oplus \cS_{[0,j]}\,,
\ee 
with
\be
	\cS_{[j,0]}=\cD(j+1,[j,0])=\cD(j+1,(j,j))\,,
	\qquad
	\cS_{[0,j]}=\cD(j+1,[0,j])=\cD(j+1,(j,-j))\,.
\ee
The character of the spin-$j$ doubleton is given by
\ba\label{short}
&&
\chi_{\cS_{[j,0]}}=\chi_{j+1,[j ,0]}-\chi_{j +2,[j -\frac12,\frac12]}
+\chi_{j +3,[j -1,0]}\,,\nn 
&&
\chi_{\cS_{[0,j]}}=\chi_{j+1,[0 ,j]}-\chi_{j +2,[\frac12,j -\frac12]}
+\chi_{j +3,[0,j -1]}\,,
\ea
in terms of the character
 $\chi_{\D,[j_+,j_-]}$ for the Verma module $\cV(\D,[j_+,j_-])$\,.
It is given by
\begin{equation}\label{long}
\chi_{\D,[j_+,j_-]}\!\left(q,x_+,x_-\right)
=q^{\Delta}\,P(q,x_+,x_-)\,\chi_{j_+}\!\left(x_+\right)\chi_{j_-}\!\left(x_-\right),
\end{equation}
where  $\chi_j$ is the $su(2)$ character in $(2j+1)$-dimensional representation:
\begin{equation}
\chi_{j}\!\left(x\right)= \frac{x^{j+\frac12}-x^{-j-\frac12}}{x^{\frac12}-x^{-\frac12}}
=\frac{\sin(j+\frac12)\a}{\sin\frac{\a}2}
\qquad [x=e^{i\,\a}]\,,
\end{equation}
and $P(q,x_+,x_-)$ takes the form,
\be
	P(q,x_+,x_-)
	=\frac1{\left(1-q\,x_+^{\frac12}\,x_-^{\frac12}\right)
	\left(1-q\,x_+^{-\frac12}\,x_-^{\frac12}\right)
	\left(1-q\,x_+^{\frac12}\,x_-^{-\frac12}\right)
	\left(1-q\,x_+^{-\frac12}\,x_-^{-\frac12}\right)}.
\ee
Therefore, the characters for $\chi_{\cS_{[j,0]}}$ and $\chi_{\cS_{[0,j]}}$ read
\ba
&& \chi_{\cS_{[j,0]}}(q,x_+,x_-)\nn
&&=q^{j+1}\,P(q,x_+,x_-)
\left[\chi_j(x_+)-q\,\chi_{j-\frac12}(x_+)\left(x_-^{\frac12}+x_-^{-\frac12}\right)
+q^2\,\chi_{j-1}(x_+)\right]\nn
&&=e^{-j\,\b}\,
\frac{\left(\cosh\b\,\cos\frac{\a_+}2-\cos\frac{\a_-}2\right)\,\csc\frac{\a_+}2\,
\sin j\a_+\,
+\sinh\b\,\cos j\,\a_+}
{2\left(\cosh\b-\cos\frac{\a_++\a_-}2\right)
\left(\cosh\b-\cos\frac{\a_+-\a_-}2\right)}\,,\nn 
&& 
\chi_{\cS_{[0,j]}}(q,x_+,x_-)=\chi_{\cS_{[j,0]}}(q,x_-,x_+)\,,
\label{spin j char}
\ea
where $q=e^{-\b}$ and $x_\pm=e^{i\,\a_\pm}$\,.
Let us remark that when \mt{j=0} and \mt{j=1/2}
the above character successfully reproduces
that of scalar and spinor doubleton
thanks to the identities $\chi_{-1/2}=0$ and $\chi_{-1}=-1$\,.
Combining $\chi_{\cS_{[j,0]}}$ and $\chi_{\cS_{[0,j]}}$\,,
we obtain the character of parity-invariant representation as
\be
	\chi^{\phantom{g}}_{\cS_j}(\b,\a_+,\a_-)
	=\chi^{\phantom{g}}_{\cS_{[j,0]}}(\b,\a_+,\a_-)+\chi^{\phantom{g}}_{\cS_{[0,j]}}(\b,\a_+,\a_-)\,.
\label{Sj char}
\ee
This character will play a key role in the subsequent analysis of this paper.

When $j$ is an integer, the boundary operator corresponding to  $\cS_j$ is the \emph{curvature} tensor, $R_{a_1b_1,\ldots,a_j b_j}$\,,
having $(j,j)$ Young diagram representation.
They are traceless --- any contraction of two indices vanish ---
and subject to the Bianchi identity
and the conservation condition,
\be
	\nabla_{[c} R_{a_1b_1],\ldots,a_\ell b_\ell}=0\,,\qquad 
	\nabla^{a_1} R_{a_1b_1,\ldots,a_\ell b_\ell}= 0\,,
	\label{conserv}
\ee
where the background is given by a conformally flat Einstein metric.
In the spinor index notation, 
the chiral and anti-chiral part of the traceless curvatures,
namely Weyl tensors, read
\be
	C_{\a_1\cdots \a_{2j}}\,,
	\qquad 
	C_{\dot\b_1\cdots \dot\b_{2j}}\,,
\ee
Here $\a_i$'s and $\dot\b_i$'s take two values and they are fully symmetric. 
The Bianch identity (or conservation condition) \eqref{conserv}
becomes  the Bargmann-Wigner equations \cite{Bargmann:1948ck},
\be
	\nabla^{\a_1\dot\b}C_{\a_1\cdots \a_{2j}}=0\,,
	\qquad 
	\nabla^{\a\dot \b_1}C_{\dot\b_1\cdots \b_{2j}}=0\,.
\ee
Even though the realization through the curvature tensor makes
manifest the conformal symmetries, 
treating the curvature as fundamental fields is not compatible
with conventional action principle.\footnote{In principle, there may exist non-conventional but manifestly conformal invariant actions involving curvature-like fields. See e.g. \cite{Francia:2002aa} for an attempt of relaxing locality to get a curvature action.}
In fact, the Bianchi identity can be solved in terms of gauge potential $\phi_{a_1\cdots a_j}$ as
\cite{Damour:1987vm,DuboisViolette:2001jk,Bekaert:2003az}.
In flat space, the solution reads
\be
    R_{a_1b_1,\ldots,a_jb_j}
    =\partial_{[a_1}\cdots \partial_{[a_j}\,\phi_{b_1]\cdots b_j]}\,,
\ee
with gauge symmetry,
\be
    \delta\,\phi_{a_1\cdots a_j}=\partial_{(a_1}\,\xi_{a_2\cdots a_j)}\,.
\ee
This makes link the curvature formulation to the Fronsdal's one \cite{Fronsdal:1978rb}
having two-derivative equation,
\be
    F_{a_1\cdots a_j}
    =\Box\,\phi_{a_1\cdots a_j} 
    -j\,\partial_{(a_1}\partial^{b}\,\phi_{a_2\cdots a_j)b}
    +\frac{j(j-1)}2\,\partial_{(a_1}\partial_{a_2}\,
    \phi_{a_3\cdots a_j)b}{}^b= 0\,.
    \label{F eq}
\ee
For the consistency, the gauge parameter and fields have to satisfy
traceless and double-traceless condition: $\xi^{a_1}{}_{a_1\cdots a_{j-1}}=0$
and $\phi^{a_1a_2}{}_{a_1\cdots a_{j}}=0$\,.
The action giving \eqref{F eq} reads
\be
    S_{\rm Fronsdal}=
    \int d^4x\,\phi^{a_1\cdots a_j}(x)\,
    G_{a_1\cdots a_j}(x)\,,
\ee
where $G_{a_1\cdots a_j}=
F_{a_1\cdots a_j}-\frac14\,s(s-1)\,\eta_{(a_1a_2}\,F_{a_3\cdots a_j)}$ 
is the spin-$j$ generalization of `Einstein tensor'.
The Fronsdal's formulation reduces to that of
Maxwell and Fierz-Pauli for the spin one and two cases.
For half-integer $j$, there exists a similar formulation making use of rank-$j$ tensor spinor \cite{Fang:1978wz}.
The Fronsdal formulation has many advantages such as more standard form of equation
and existence of conventional action principle.
However, on the contrary to the curvature formulation,
it does not admit any local form of conformal transformation at the level of gauge potential $\phi_{a_1\cdots a_j}$
(see \cite{Barnich2015} and references therein for more details).
Moreover, for spin $j\ge3/2$\,, it can be defined only
on a conformally flat and Einstein background 
\cite{Christensen:1978md,Aragone:1979hx}.

\subsection{Single-Trace Operators of the Vector Model CFT}

\subsubsection*{U(N) Vector Model}

The spin-$j$ $U(N)$ model is
described by $N$ copies of complex massless spin-$j$ fields.
In Fronsdal formulation, the spin-$j$ action is given by
\be
    S_{\rm CFT}=\int d^4x\,\sum_{i=1}^N \bar \phi_i^{a_1\cdots a_j}\,
    G_{i\,a_1\cdots a_j}\,.
\ee
All $U(N)$-invariant single-trace operators are given by
bilinear in the Weyl curvatures $C_{i\,\a_1\cdots \a_{2j}}$ and 
$C_{i\,\b_1\cdots \b_{2j}}$, and their complex conjugates.
The operator content can be identified group theoretically and divided into three parts as
\be
	\cH_{j,U(N)}=\cH^{\rm\sst Sym}_{j,U(N)}\oplus \cH^{\rm\sst MixSym}_{j,U(N)}
	\oplus\cH^{\rm\sst Massive}_{j,U(N)}
	\,.
\ee
The first part corresponds the cross product
(in considering the fact that $\cS_j=\cS_{[j,0]}\oplus \cS_{[0,j]}$),
\be
    \cH^{\rm\sst Sym}_{j,U(N)}= (\cS_{[j,0]}\otimes \cS_{[0,j]})
    \oplus (\cS_{[0,j]}\otimes \cS_{[j,0]}) =
    2 \bigoplus_{s=2j}^\infty 
	\cD(s+2,[\tfrac s2,\tfrac s2])\,,
	\label{sym cur}
\ee
and the second and third parts come from the sum of `square',
\be 
    \cH^{\rm\sst MixSym}_{j,U(N)}
    \oplus \cH^{\rm\sst Massive}_{j,U(N)}
    =
    (\cS_{[j,0]}\otimes \cS_{[j,0]})
    \oplus (\cS_{[0,j]}\otimes \cS_{[0,j]})\,, 
    \label{FF long}
\ee 
with
\be
     \cH^{\rm\sst MixSym}_{j,U(N)}=
     	\bigoplus_{s=2j+1}^\infty
	\!\cD(s+2,[\tfrac s2+j,\tfrac s2-j]_{\rm\sst PI})\,,
	\label{FF j}
\end{equation}
and
\be
     \cH^{\rm\sst Massive}_{j,U(N)}
    =2\,\cD(2j+2,[0,0])\oplus
	\bigoplus_{r=1}^{2j}
	\cD(2j+2,[r,0]_{\rm\sst PI})\,.
\ee
Here, the subscript ${\rm\sst PI}$ means the parity-invariant combination,
\be
	[j_+,j_-]_{\rm\sst PI}=[j_+,j_-]\oplus [j_-,j_+]\,.
\ee
The UIRs  in $\cH^{\rm\sst Massive}_{j,U(N)}$ 
\eqref{FF long} are long
and correspond to non-conserved currents (dual to massive fields in AdS).
Their explicit forms are given by
\ba
    \cO_{\a_1\cdots \a_{2r}}
    \eq C_{(\a_1\cdots\a_r|\g_{1}\cdots\g_{2j-r}}\,
    \bar C_{\a_{r+1}\cdots \a_{2r})}{}^{\g_{1}\cdots\g_{2j-r}}\,,
    \nn 
    \cO_{\dot\b_1\cdots \dot\b_{2r}}
    \eq C_{(\dot\b_1\cdots\dot\b_r|\dot\d_{1}\cdots\dot\d_{2j-r}}\,
    \bar C_{\dot\b_{r+1}\cdots \dot\b_{2r})}{}^{\dot\d_{1}\cdots\dot\d_{2j-r}}\,,
    \label{O r}
\ea
Here, the summation over the internal $U(N)$ index should be understood.
The UIRs of $\cH^{\rm\sst Sym}_{j,U(N)}$ \eqref{sym cur} are the symmetric rank $s\ge 2j$ conserved currents
which are nothing but the traceless Bel-Robinson current \cite{Berends:1985xx,Gelfond:2006be},
\ba 
    J_{\a_1\cdots \a_s}{}^{\dot\b_1\cdots \dot \b_s}
    \eq  \sum_{n=0}^{s-2j}(-1)^n\binom{s}{n}\,\binom{s}{2j+n}
    \times \nn 
    &&\times\, 
    \partial_{(\a_1}^{(\dot\b_1}\cdots  
    \partial_{\a_n\phantom{|}}^{\dot\b_n\phantom{|}}
    C^{\phantom{|}}_{\a_{n+1}\cdots \a_{n+2j}\phantom{|}}\,
    \partial_{\a_{2j+n+1}\phantom{|}}^{\dot\b_{n+1}\phantom{|}}\cdots  
    \partial_{\a_{s})}^{\dot\b_{s-2j}\phantom{|}}\bar 
    C^{\dot\b_{s-2j+1}\cdots \dot\b_s)}_{\phantom{|}}\,.
\ea 
The other symmetric conserved currents $J'_{\a_1\cdots \a_s\dot\b_1\cdots \dot \b_s}$
have the same form as above but $C$ and
$\bar C$ interchanged.
Lastly, the UIRs in $\cH^{\rm\sst Sym}_{j,U(N)}$ 
\eqref{FF j} are the mixed-symmetry conserved currents 
having the form,
\ba
    &&J_{\a_1\cdots \a_{s+2j}}{}^{\dot\b_1\cdots \dot \b_{s-2j}}
  	=  \sum_{n=0}^{s-2j} A^{s,j}_{n}
    \times \nn 
    &&\qquad \times\, 
    \partial_{(\a_1}^{(\dot\b_1}\cdots  
    \partial_{\a_n\phantom{|}}^{\dot\b_n\phantom{|}}
    C^{\phantom{|}}_{\a_{n+1}\cdots \a_{n+2j}\phantom{|}}\,
    \partial_{\a_{2j+n+1}\phantom{|}}^{\dot\b_{n+1}\phantom{|}}\cdots  
    \partial_{\a_{s}\phantom{|}}^{\dot\b_{s-2j})\phantom{|}}
    \bar C_{\a_{s+1}\cdots \a_{s+2j})}^{\phantom{|}}\,.
\ea
with $s\ge 2j+1$ and the coefficient $A^{s,j}_n$ to be determined by 
the conservation condition,
\be
    \partial^{\a_1\dot\b_1} J_{\a_1\cdots \a_{s+2j}\dot\b_1\cdots \dot \b_{s-2j}}=0\,.
\ee
The explicit derivation of these currents can be
most conveniently
done in the unfolded formulation.
See \cite{Gelfond:2003vh,Gelfond:2006be,Gelfond:2008ur}
 for the details.

\subsubsection*{O(N) Vector Model}

When the massless spin-$j$ fields are real, 
the model becomes $O(N)$, and its single-trace operator spectrum is given by
\be
	\cH_{j,O(N)}=
	\cS_j\otimes_{\rm cyc} \cS_j
	=\cH_{j,O(N)}^{\rm\sst Sym}
	\oplus \cH_{j,O(N)}^{\rm\sst MixSym}
		\oplus \cH_{j,O(N)}^{\rm\sst Massive}
	\,, 
\ee	
where $\otimes_{\rm cyc}$ is the (anti-)symmetric tensor product
for (half-)integer $j$ and 
\ba	
	\cH_{j,O(N)}^{\rm\sst Sym}\eq 
  	\bigoplus_{s=2j}^\infty 
	\cD(s+2,[\tfrac s2,\tfrac s2])\,,\nn 
	\cH_{j,O(N)}^{\rm\sst MixSym}\eq 
		\bigoplus_{{\rm even}\,s\ge 2j+1}
	\cD(s+2,[\tfrac s2+j,\tfrac s2-j]_{\rm\sst PI})\,,\nn
	\cH_{j,O(N)}^{\rm\sst Massive}\eq
	2\,\cD(2j+2,[0,0])\oplus\!
	\bigoplus_{2\le {\rm even}\,r\le 2j}\!
	\cD(2j+2,[r,0]_{\rm\sst PI})\,.
\ea
Correspondingly, the operators $\cO_{\a_1\cdots \a_{2r}}$ 
with odd $r$
and the symmetric currents $J'_{\a_1\cdots \a_s\dot \b_1\cdots \dot\b_s}$ are projected out.
Let us note that in both of $U(N)$ and $O(N)$
model, they do not admit any conserved rank-two tensor if $j\ge 3/2$\,.
Hence, there is no energy-momentum tensor in these models,
implying that the AdS dual theory --- type-$j$ higher-spin theory ---
is non-gravitational.

\section{Classical Type-$j$ Higher Spin Theory in AdS$_5$}
\label{sec:classicalHS}

In this section, we consider 
the AdS$_5$ dual of the spin-$j$ vector model CFT in four dimensions.
They will be referred here as to type-$j$ higher-spin theory.
Starting from the discussion of their field content,
we discuss about its underlying higher-spin algebra.  

\subsection{Field Content}

The field content of type-$j$ higher-spin theories 
are defined to coincide with the single-trace operator spectrum of the dual theory,
which have been reviewed in the previous section.
Depending on whether the boundary theory 
is $U(N)$ or $O(N)$, the bulk theory differs:
non-minimal theory dual to the former
and minimal one dual to the latter.

There appear three types of fields. The first one is the massive fields
\be
	\Pi_{\mu_1\cdots \mu_r,\n_1\cdots \n_r}
	\quad [r=0,(1),\ldots,(2j-1),2j]\,,
	\qquad \Pi'\,,
\ee	
where minimal theory has only even $r$
and there are two scalar fields $\Pi$ and $\Pi'$
with the same mass.
The above fields are
dual to the operator $\cO_{\a_1\cdots \a_{2r}}$ and its conjugate \eqref{O r},
hence carrying $\cD(2j+2,(r,r)_{\rm\sst PI})$ representation. 
Their mass-squared value 
is given by
\be
	m^2=(2j+2)(2j-2)-2r\,.
\ee
These fields are the generalization of the scalar field in the spectrum of type A theory.
Next, we have two types of  massless fields:
the symmetric one and the mixed-symmetry one
\be 
    \begin{split}
    \Phi_{\mu_1\cdots \mu_s}\,,
    \quad (\Phi'_{\mu_1\cdots \mu_s})
	\quad [s=2j,2j+1,\ldots]\,,\\
	\Psi_{\mu_1\cdots\mu_s,\nu_1\cdots\nu_{2j}}
	\quad [s=(2j+1),2j+2,\ldots]\,.
	\end{split}
\ee 
In the minimal theory, we have only one copy
of symmetric fields and even $s$ mixed symmetry fields.
The above fields are all gauge fields,
hence the theory has
infinite amount of gauge symmetries analogously
to the type A, B, C models.
The gauge symmetry takes the form of
\ba
	&& \delta\,\Phi_{\mu_1\cdots \mu_s}
	= \nabla_{(\mu_1}\varepsilon_{\mu_2\cdots \mu_s)}+
	\cO(g)\,,\nn
	&& \delta\,\Psi_{\mu_1\cdots \mu_s,\nu_1\cdots \nu_{2j}}
	= \nabla_{(\mu_1}\xi_{\mu_2\cdots \mu_s),\mu_1\cdots \mu_{2j}}
	+\cO(g)\,,\nn
	&& \delta\,\Pi_{\mu_1\cdots \mu_r,\n_1\cdots \n_r}=\cO(g)\,,
	\label{gauge}
\ea
where $\nabla_\mu$ is the AdS covariant derivative
and $g$ is the coupling constant of the bulk theory,
hence $\cO(g)$ contains the field-dependent terms.

Apart from the type A theory, in all cases of type-$j$, the theory involves  mixed-symmetry gauge fields $(s,2j)_{\rm\sst PI}$.
The degrees of freedom (DoF) of  mixed-symmetry fields with definite parity --- $(s,2j)$ or $(s,-2j)$ --- can be easily counted as
\be
	{\rm dim}\left(\pi^{O(4)}_{(s,\pm2j)}\right)-{\rm dim}\left(\pi^{O(4)}_{(s-1,\pm2j)}\right)
	=2\,s+1\,,
\ee
where $\pi^{O(4)}_{(\ell_1,\ell_2)}$ indicates the $O(4)$ tensor representation corresponding 
to the Young diagram $(\ell_1,\ell_2)$\,.
Let us note that the DoF do not depend on the value of $j$ (of type-$j$ theory). 
Hence, all the mixed-symmetry gauge fields of definite parity
have exactly same number of DoF, and in particular coincides 
with the totally-symmetric field.
In terms of parity-invariant fields carrying $\cD(s+2,(s,2j)_{\rm\sst PI})$, the mixed-symmetry fields with $s\ge 2j+1$ have twice many DoF as the symmetric field. 
In the flat limit \cite{Brink:2000ag,Boulanger:2008up,Boulanger:2008kw,Alkalaev:2009vm,Alkalaev:2011zv}, the mixed-symmetry gauge field of $(\ell_1,\ell_2)$ type decomposes into the massless helicity modes
corresponding to the $O(3)$ --- massless little group in five dimensions --- Young diagram,
\be 
\parbox{83pt}{
	\begin{tikzpicture}
	\draw (0,0) rectangle (2.8,0.4);
	\node at (1.4,0.2){${\st \ell_1}$};
	\end{tikzpicture}}\oplus\ 
\parbox{83pt}{
	\begin{tikzpicture}
	\draw (0,0) rectangle (0.4,0.4);
	\draw (0,0.4) -- (0,0.8) -- (2.8,0.8) -- (2.8,0.4) -- (0.4,0.4);
	\node at (1.4,0.6){${\st \ell_1}$};
	\end{tikzpicture}}\oplus\ 
	\parbox{83pt}{
	\begin{tikzpicture}
	\draw (0,0) rectangle (0.4,0.4);
	\draw (0.4,0) -- (0.8,0) -- (0.8,0.4); 
	\draw (0,0.4) -- (0,0.8) -- (2.8,0.8) -- (2.8,0.4) -- (0.4,0.4);
	\node at (1.4,0.6){${\st \ell_1}$};
	\end{tikzpicture}}\oplus\ 
	\cdots\ 
	\oplus\ 
	\parbox{83pt}{
	\begin{tikzpicture}
	\draw (0,0) rectangle (2.4,0.4);
	\draw (0,0.4) -- (0,0.8) -- (2.8,0.8) -- (2.8,0.4) -- (2.4,0.4);
	\node at (1.4,0.6){${\st \ell_1}$};
	\node at (1.2,0.2){${\st \ell_2}$};
	\end{tikzpicture}}.	
\ee 
Besides the first two helicity modes, all the rest having more than one boxes in the second row vanish identically hence do not propagate in five dimensions.\footnote{However, they may become relevant for the would-be gauge modes surviving on the boundary.}
The first helicity mode is that of totally symmetric field
and the second one can be dualized to give again totally symmetric DoF:
\be 
\parbox{83pt}{
	\begin{tikzpicture}
	\draw (0,0) rectangle (0.4,0.4);
	\draw (0,0.4) -- (0,0.8) -- (2.8,0.8) -- (2.8,0.4) -- (0.4,0.4);
	\node at (1.4,0.6){${\st \ell_1}$};
	\end{tikzpicture}}\sim\ 
\parbox{83pt}{
	\begin{tikzpicture}
	\draw (0,0) rectangle (2.8,0.4);
	\node at (1.4,0.2){${\st \ell_1}$};
	\end{tikzpicture}}\,.
\ee
This accounts the twice many DoF of the mixed-symmetry gauge fields in AdS$_5$\,.
Even though their DoF are related to those of the totally-symmetric field and they reduce
to the latter in the flat limit,
these mixed-symmetry gauge fields are genuinely different representations in AdS$_5$.

\subsection{Cubic Interactions}

Let us consider the interaction nature of the type-$j$ theory. Like the field content, the interaction structures
are to match with the correlation functions of boundary operators through the Witten diagram. 
We shall mostly focus on the cubic interactions of these higher spin theories which
are dual to the three-point correlators on the boundary.
Cubic interactions of a gauge theory already tells many characteristic properties of the theory. 
The appendix C.1 of \cite{Gunaydin:2016amv} contains several discussions on the structure of cubic interactions.
Here, we provide additional discussions which, we hope, help to better understand the theory.

To make more clear our discussion, let us begin with a few comments on the generality.
When a gauge theory has cubic interactions, then the 
field-dependent part of the gauge 
transformation (the $\cO(g)$ part of \eqref{gauge}) should have 
terms linear in the field, so that the latter
compensate the linearized gauge transformation of cubic vertices.
Cubic interactions compatible with gauge symmetries can be classified into three different groups. The first group is what is called non-Abelian interactions, which is the part encoding the information of the underlying global symmetry which 
the theory is gauging. In other words, the bracket of the global symmetry generators are determined by these cubic vertices hence in the absence of these interactions, the generators would commute hence
the algebra would remain Abelian. Typical such interactions are the
self-couplings of Einstein gravity and Yang-Mills theory.
Abelian cubic interactions are split again into two groups. The first one is `deforming' interactions
which requires that the 
gauge transformation of the relevant fields 
have linear term in field. 
Typical examples are the minimal coupling to the matter. Due to this coupling, matter field transforms under the gauge symmetry. The rest of the couplings are `non-deforming' interactions
whose presence does not necessitate any
linear term in the gauge transformations.
Non-minimal curvature interactions are the typical examples.

\subsection*{Higher Spin Algebra}

Now considering back the type-$j$ theory, 
all its non-Abelian interactions are in fact to be dictated by global symmetry, that is, higher spin algebra.
With additional input that the boundary dual theory is the massless spin-$j$, we can identify the relevant higher spin algebra. It has two types of generators,
which are respectively the
solutions of the Killing equations,
\be
    \nabla_{(\mu_1}\varepsilon_{\mu_2\cdots \mu_s)}=0\,,
    \qquad 
    \nabla_{(\mu_1}\xi_{\mu_2\cdots \mu_s),\mu_1\cdots \mu_{2j}}=0\,.
\ee
As $O(2,4)$ tensors, they are characterized by
\be
    \parbox{105pt}{
	\begin{tikzpicture}
	\draw (0,0) rectangle (3.6,0.8);
	\draw (0,0.4) -- (3.6,0.4);
	\node at (1.8,0.6){$s-1$};
	\node at (1.8,0.2){$s-1$};
	\end{tikzpicture}}
	\quad [s\ge 2j]\,,
	\qquad 
	\parbox{105pt}{
	\begin{tikzpicture}
	\draw (0,0) rectangle (3.6,0.8);
	\draw (0,0.4) -- (3.6,0.4);
	\node at (1.8,0.6){$s-1$};
	\node at (1.8,0.2){$s-1$};
	\draw (0,0) -- (0,-0.4) -- (2,-0.4) -- (2,0);
	\node at (1,-0.2){$2\,j$};
	\end{tikzpicture}}
	\quad [s\ge 2j+1]\,,
\ee
and again depending on 
whether the theory is minimal (or not), 
the algebra contains 
one (or two) copy of $(s-1,s-1)$ generators
and even $s$ (and odd $s$)
generators of mixed-symmetry type $(s-1,s-1,2j)$.

One of the simplest way to understand  the type-$j$ higher spin algebra,
which we shall refer as to $HS_j$, is
to view it as the maximal symmetry
of the boundary theory,
that is, the endomorphism algebra
of the spin-$j$ doubleton Hilbert space:
\be
    HS_j={\rm End}(\cS_j)\,.
\ee
Since $\cS_j$ can be decomposed into $\cS_j=\cS_{[j,0]}\oplus \cS_{[0,j]}$
where $\cS_{[j,0]}=\cD(j+1,[j,0])$\,,
the type-$j$ higher spin algebra also
splits into
\be
    HS_j={\rm End}(\cS_{[j,0]})\oplus {\rm End}(\cS_{[0,j]})
    \oplus {\rm Hom}(\cS_{[j,0]},\cS_{[0,j]})
    \oplus {\rm Hom}(\cS_{[0,j]},\cS_{[j,0]})\,.
\ee
The first two parts
${\rm End}(\cS_{[j,0]})\oplus {\rm End}(\cS_{[0,j]})$
form an subalgebra,
and they are isomorphic to each other,
\be
    hs_j={\rm End}(\cS_{[j,0]})\simeq {\rm End}(\cS_{[0,j]})\,.
    \label{hs j}
\ee
In the minimal theory case, we have only one copy of $hs_j$
as subalgebra. $hs_j$ contains only
the generators of symmetric spin $s\ge 2j$
and can be obtained 
from the universal enveloping algebra of $su(2,2)$ by quotienting it with 
the annihilator of spin-$j$ module.\footnote{The annihilator of spin-$j$ module is the maximal ideal of the universal enveloping algebra of $su(2,2)$, namely Joseph ideal \cite{Joseph:1974hr}. 
The representations underlying Joseph ideal
are called minimal representations,
so the spin-$j$ doubletons
are the minimal representations of $su(2,2)$.
Joseph ideals and minimal representations
are extensively studied in mathematics community,
and its discussion in the context of higher spin algebra can be found in \cite{Joung:2014qya}.}
At the level of Lie algebra,
one can consider $j$ as a continuous parameter,
say $\l$,
then $hs_j$ can be enhanced to $hs_\l(su(2,2))$
containing all the generators of symmetric spin $s\ge 1$\,.
More explicitly, it is given as the coset
of the tensor algebra generated by $su(2,2)$ generators
by the equivalence relation \cite{Fronsdal:2009},
\be
    L^{[a}_{\ b}\otimes L^{c]}_{\ d}
    +\d^{[a}_{(b}\,L^{c]}_{\ d)}
  	+\l\, \d^{[a}_{[b}\,L^{c]}_{\ d]}
  	+\frac{\l^2-1}{4}\, \d^{[a}_{[b}\,\d^{c]}_{d]}\sim 0\,,
    \label{L rel}
\ee
where $L^a{}_b$ are the $su(2,2)$ generators.
When the parameter $\l$ takes an half-integer values $j$
(so that the Lie-algebra representation uplifts to that of Lie-group),
the algebra $hs_j(su(2,2))$ develops the ideal algebra $hs_j$ \eqref{hs j}. The coset algebra $hs_j(su(2,2))/hs_j$ is isomorphic to
\be
    u\!\left(\frac23\,j\left(j^2+\frac12\right),
    \frac23\,j\left(j^2-1\right)\right)
    \quad
    {\rm or}
    \quad 
     u\!\left(\frac23\,j\left(j^2-\frac14\right),
    \frac23\,j\left(j^2-\frac14\right)\right),
\ee 
depending on whether $2j$ is even or odd.\footnote{In fact, the
algebra $hs_\l(su(2,2))$ is just a particular case of $sl_N$ higher spin algebra: in the same way one can define $hs_\l(sl_N)$ and show that the appearance of ideal and finite quotient algebra is a generic feature of this series of higher spin algebra:
 when $N(\l-1)/2$ takes an integer value $M$,
$hs_\l(sl_N)$ develops an ideal consisting of generators with $r\ge M$ 
and the corresponding coset algebra becomes $gl_{\binom{N+M-1}{M}}$
consisting of the generators with $r=0,1,\ldots,M-1$\,.
The most well-known example of this would be the $N=2$ case where we get
the half of the three dimensional higher spin algebra, often referred to as $hs(\l)$\,.
Let us also remind the reader that
the coset algebras appearing in this way starting from the right real form $su(1,1)$ and $su(2,2)$
respectively for three and six dimensional cases do not have the right reality structure to be interpreted
as unitary theory of massless higher spin fields because they have alternating kinetic term sign.
Instead the ideal parts have the right reality structure: its bilinear form has all same sign for each spin blocks
(of course, in each spin block there are negative norm components, but what is important is the relative sign between different spins).}
The one-parameter family
algebra has been first considered in
higher spin context in \cite{Fradkin:1989yd}, and more recent discussions 
can be found in \cite{Boulanger:2011se,Manvelyan:2013oua, Govil:2013uta,Boulanger:2013zza,Joung:2014qya}.

The type-$j$ higher spin algebra $HS_j$ can be  realized by oscillators with commutation relation,
\be
    [\,Y_A^\a\,,\,Y_B^\b\,]_\star=\eta_{AB}\,\e^{\a\b}\,,
    \qquad 
    \{\,\th_A^i\,,\,\th_B^j\,\}_\star
    =\eta_{AB}\,\delta^{ij}\,,
\ee 
where $A,B$, $\a,\b$ and $i,j$
are respectively the $O(2,4)$, 
$Sp(2)$ and $O(2j)$ fundamental indices.
Noticing that the $so(2,4)$, realized in these oscillators as
\be
    M_{AB}=Y_A^\a\,Y_{B\b}+\theta_A^i\,\th_{Bj}\,,
\ee
commutes with the $osp(2j|2)$,
\be
    K^{\a\b}=Y_A^\a\,Y^{A\b}\,,
    \qquad 
    R^{ij}=\th_A^i\,\th^{Aj}\,,
    \qquad 
    S^{\a i}=Y_A^\a\,\th^{Ai}\,,
\ee
one can define a higher spin algebra
from the Weyl-Clifford algebra $A_{6,12j}$ freely generated by $Y_A^a$ and $\th_A^i$.
More precisely, it is given by
the quotient of the $osp(2j|2)$-centralizer
by the $osp(2j|2)$ ideal. 
See \cite{Vasiliev:2004cm,Bekaert:2009fg} for more details.
This realization makes use of vector oscillators,
but for $so(2,4)$ it is more convenient
to use spinor oscillators based 
on $su(2,2)$ (see \cite{Sezgin:2001zs,Beisert:2003te,Beisert:2004di,Vasiliev:2004cm} for related discussions)
with commutation relations,
\be 
    [\,y^a\,,\bar y_b\,]_\star =\delta^{a}_b\,,
    \qquad 
    \{\,\th^i_a\,,\,\bar\th^b_j\,\}_\star
    =\delta^i_j\,.
\ee
The $a,b$ are $i,j$ now the $SU(2,2)$ and 
$O(2j)$ fundamental indices.
With these oscillators, the $su(2,2)$ generators are given by
\be 
    L^{a}{}_b=y^a\,\bar y_b+\bar\theta^a_i\,\theta_b^i\,,
\ee
and they commute with $gl(2j|1)$ given by
\be
    K=y^a\,\bar y_b\,,
    \qquad 
    R^{i}{}_{j}=\bar\theta_a^i\,\theta^a_i\,,
    \qquad 
    S^{i}=y^a\,\theta_a^i\,
    \qquad
    \bar S_i=\bar y_a\,\bar\theta^a_i\,.
\ee
Hence, one can obtain a higher spin algebra 
as the $gl(2j|1)$-centralizer
of Weyl-Clifford algebra $A_{4,8j}$
by the $gl(2j|1)$ ideal.
However the higher spin algebras defined in these ways are not the type-$j$ one, $HS_j$\,,
because the former 
always include the $su(2,2)$ generators 
associated with the graviton in the bulk
whereas the type-$j$ theory
does not have any symmetric spin fields lower than spin $2j$.
Though not manifest in the oscillator form, 
these oscillator algebras are not semi-simple
but contains an ideal corresponding to $HS_j$\,.  
By focusing on the subalgebra $hs_j$ sector,
the oscillator realization corresponds
hence to $hs_j(su(2,2))$ rather than 
the ideal $hs_j$\,.
See the forthcoming work \cite{EuihunKarapet}
for the detailed analysis.

\section{One-Loop Partition Function of Type-$j$ Theory}
\label{sec:oneloopHS}

The partition function of a generic theory in AdS$_5$
can be expanded order by order in the number of loops as 
\be
\Gamma_{\cH}= -\log\!\left[
\int \prod_i\cD\varphi_i
\,\exp\!\left(-{1\over g}\,S_{\cH}[\varphi]\right)\right] 
	={1\over g}\,S_{\cH}+\G^{\sst (1)}_{\cH}+\cO(g)\,,
\ee
where the first term, $S_{\cH}$, is the classical action evaluated with vacuum field configuration
and  the one-loop part $\G^{\sst (1)}_{\cH}$ can be computed by evaluating the Gaussian path integral.
Since $\G^{\sst (1)}_{\cH}$ depends only on the field content $\cH$\,,
we can calculate the quantity for the individual fields and resum them for the total one-loop result.
In \cite{Giombi:2013fka, Giombi:2014iua}, this resummation has been carried out
for Vasiliev theory on AdS$_{d+1}$ with $S^d$ boundary. 
The resummation requires additional regularization
but it turns out that if we keep the UV regulator while resumming then the summation is convergent
and the UV divergence also cancels in all the cases.
Recently, in \cite{Bae:2016rgm,Bae:2016hfy}, the authors of the current paper have introduced
a method of doing this resummation with regulator turned on. In other words, it allows to 
calculate the spectral zeta function directly from the character underlying the field content of the theory.
This method, referred to as Character Integral Representation of Zeta function (CIRZ), 
allows us to bypass the explicit identification of the field content and 
the subtlety arising in the resummation process.

\subsection{One-Loop Vacuum Energy in AdS$_5$ with S$^4$ Boundary}

The renormalized one-loop partition function (or free energy) will be
referred to as \emph{vacuum energy} and takes the form,
\be
\Gamma^{\sst (1)\,\rm ren}_{\cH} =(-1)^F\,\g_{\cH}\,\log R\,,
\ee
where $R$ is the IR cut-off of the AdS$_5$ space,
$\g_{\cH}$ 
is a constant which depends on the theory
and $F$ is 0 for bosonic $\cH$ and 1 for fermionic $\cH$.
Due to the $\log R$ term, for the entire AdS$_5$ with $R\to \infty$,
the above quantity is divergent.
Nevertheless, the way it diverges, namely the factor $\g_{\cH}$, encodes
the one-loop information of the theory.
Using the CIRZ method \cite{Bae:2016hfy,Bae:2016rgm}, we have shown that 
 $\g_{\cH}$ is given by the sum of three quantities,
\begin{equation}
\g_{\cH}=\gamma_{\mathcal{H}|2}+\gamma_{\mathcal{H}|1}+\gamma_{\mathcal{H}|0}\,,
\label{g H}
\end{equation}
where each of $\gamma_{\mathcal{H}|n}$ is the contour integral,
\begin{equation}\label{gammahn}
\gamma_{\mathcal{H}|n} = -\left(-4\right)^n n! \oint {d\beta\over 2\pi i}\,{f_{\mathcal{H}|n}(\b)\over\beta^{2\left(n+1\right)}}\,,
\end{equation}
of a function given by the character of the theory:
\begin{equation}\label{f H}
\begin{split}
f_{\mathcal{H}|2}(\b)&= {\sinh^4{\tfrac\beta2}\over 2}\,\chi_{\mathcal{H}}\left(\beta,0,0\right),\\
f_{\mathcal{H}|1}(\b)&= \sinh^2{\tfrac\beta2}\left[{\sinh^2{\tfrac\beta2}\over 3}-1-\sinh^2{\tfrac\beta2}\left(\partial_{\alpha_1}^2 +\partial_{\alpha_2}^2\right)\right]\chi_{\mathcal{H}}\left(\beta,\alpha_{+},\alpha_{-}\right)\bigg|_{\alpha_{\pm}=0},\\
f_{\mathcal{H}|0}(\b)&=\left[1 + {\sinh^2\tfrac\beta2\left(3-\sinh^2\tfrac\beta2\right)\over 3} \left(\partial_{\alpha_1}^2 +\partial_{\alpha_2}^2\right)\right.\\&\qquad\left. -{\sinh^4\tfrac\beta2\over 3}\left(\partial_{\alpha_1}^4-12\,\partial_{\alpha_1}^2\partial_{\alpha_2}^2+\partial_{\alpha_2}^4\right)\right]\chi_{\mathcal{H}}\left(\beta,\alpha_{+},\alpha_{-}\right)\bigg|_{\alpha_{\pm}=0}.
\end{split}
\end{equation}
In the AdS theories dual to vector models on the boundary,
 the functions $f_{\cH|n}$ are analytic except for poles at the origin.
In such cases, we can take the contour as the circle around the origin and use the residue theorem.
In the end, the quantities $\g_{\cH|n}$ are simply proportional to the
$\b^{2n+1}$ Laurent coefficient  of $f_{\cH|n}$\,.
Armed with the results
(\ref{g H}\,--\,\ref{f H}), let us calculate the 
vacuum energy of the non-minimal and minimal 
type-$j$ higher spin theory in  AdS$_5$\,.

In the type A, B and C cases, the vacuum energies
were found to be proportional to those
of spin 0,1/2 and 1 doubletons treated as if they are bulk fields.
In fact, these quantities also happen to coincide with
the genuine  one-loop partition function of boundary theory on $S^4$ 
once the standard UV/IR correspondence of AdS/CFT is used.
Having in mind that this might generalize to the type-$j$ theories,
we first calculate the vacuum energy of spin-$j$ doubleton
treated as if it is AdS$_5$ field.
Eventually, the latter quantity will be related to the vacuum energy of type-$j$ theory.

\subsubsection*{Vacuum Energy of Spin-$j$ Doubleton}

We first calculate the AdS$_5$ vacuum energy of spin-$j$ doubleton
even though it can be better described as massless spin-$j$ field on $S^4$\,.
In principle, we would need a AdS$_5$ description of spin-$j$ doubleton 
to compute such a quantity,
but the CIRZ method allows us to avoid this step.
Practically, it is sufficient to know the character $\chi_{\cS_j}$ of spin-$j$ doubleton representation,
which is already given in \eqref{Sj char}.
We may therefore calculate first the functions $f_{\cS_j|n}$
following the recipe \eqref{f H} to obtain
\begin{equation}
\begin{split}
f_{\cS_j|2}(\b) &= \frac18\,e^{-j\b}
\left[2\,j\,(\cosh\b-1)+\sinh\b\right],\\
f_{\cS_j|1}(\b) &= \frac{1}{12} \,e^{-j\b}
\left[4\left(j^3+j\right)(\cosh\b-1)+
\left(6\,j^2+1\right)\sinh\b\right],\\
f_{\cS_j|0}(\b) &=\frac16\,e^{-j\b}\,j
\left[ 2\,j^2(j^2+2)\cosh\b+(5\,j^3+j)\,\sinh\b-2(j^2-1)^2\right].
\end{split}
\end{equation}
We then simply Laurent expand the above to extract
the relevant coefficient to get
\be
\gamma_{\cS_j|2} = \frac{15\,j^4-1}{30}\,,\qquad 
\gamma_{\cS_j|1} = \frac{6\,j^4-3\,j^2+1}{18}\,, \qquad 
\gamma_{\cS_j|0} =\frac{j^4-j^2}{2} \,.
\label{gamma S j}
\ee
Finally, summing these three numbers, we obtain the vacuum energy as
\begin{equation}\label{vacjsingleton}
\Gamma^{\sst (1)\,\text{ren}}_{\cS_j} = \left(-1\right)^{2j}\,\frac{60\,j^4-30\,j^2+1}{45}\,\log R\,,
\end{equation}
where the overall $\left(-1\right)^{2j}$ accounts for fermionic statistics of particles with half-integer $j$.
We readily see that these expressions reduce to their previously computed counterparts for the special case of $j=0,1/2$ and 1:
\begin{equation}
\Gamma^{\sst (1)\,\text{ren}}_{\cS_0} = \frac{1}{45}\,\log R\,,
\qquad
\Gamma^{\sst (1)\,\text{ren}}_{\cS_{\frac12}} = \frac{11}{180}\,\log R\,,
\qquad
\Gamma^{\sst (1)\,\text{ren}}_{\cS_1} = \frac{31}{45}\,\log R\,.
\end{equation}
As an side, the vacuum energies of the chiral and anti-chiral singletons $\cS_{[j,0]}$ and $\cS_{[0,j]}$ are
the half of the vacuum energy of $\cS_j$\,.
We now turn to the holographic duals of vector models built from such boundary fields.

\subsubsection*{Non-Minimal Type-$j$ Theory}

This theory is the AdS$_5$ dual of the $U(N)$ vector model built from the complex spin-$j$ doubletons. Again, to compute the one-loop vacuum energy by the CIRZ method, it is sufficient to identify the underlying character.
For  the non-minimal case, it is  given by
\begin{equation}
\chi_{j,\text{non-min}}\!\left(\beta,\alpha_1,\alpha_2\right) = \chi_{\cS_j}\!\left(\beta,\alpha_1,\alpha_2\right)^2\,.
\label{non min char}
\end{equation}
We begin with the case of vector models built from the the parity invariant singleton. The expressions for the $f_{j,\text{non-min}|n}$ are quite lengthy and are hence omitted here. We directly write the expressions for the $\gamma_{j,\text{non-min}|n}$ defined in \eqref{gammahn} as
\begin{equation}
\begin{split}
\gamma_{j,\text{non-min}|2} &= \frac{2}{105}\,n_j \left(72\,j^4-24\,j^2+1\right),\\
\gamma_{j,\text{non-min}|1} &= {4\over 315}\,n_j\left(60\,j^4-27\,j^2+2\right),\\
\gamma_{j,\text{non-min}|0} &=\frac{8}{15}\,
n_j\,(j^4-j^2)\,,
\end{split}
\label{gamma type j}
\end{equation}
where $n_j$ is an integer given by
\be
    n_j=\frac{\left(2\,j-1\right)2\,j\left(2\,j+1\right)}6\,.
\ee
These may in turn be summed up to give
\be
\Gamma^{\sst (1)\,\text{ren}}_{j,\text{non-min}} 
= 
\frac{2}{45}\,n_j\,\left(60\,j^4-30\, j^2+1\right)\log R=
\left(-1\right)^{2j}\,
n_j\,2\,\Gamma^{\sst (1)\,\text{ren}}_{\cS_j}\,,
\label{1lvacjnonmin}
\ee
where we have used the expression \eqref{vacjsingleton} in the second step.
One can see that the vacuum energy vanishes for $j=0,1/2$ reproducing the result of type A and B \cite{Giombi:2014iua}. Further, for $j=1$ we obtain the relation,
\begin{equation}
\Gamma^{\sst (1)\,\text{ren}}_{j=1,\text{non-min}} =2\, \Gamma^{\sst (1)\,\text{ren}}_{\cS_1},
\end{equation}
which reproduces the result of \cite{Beccaria:2014zma}. 

Let us note that the vacuum energy of the non-minimal theory can be split into the contribution
of symmetric fields in $\cH^{\rm\sst Sym}_{j,\text{non-min}}$ \eqref{sym cur}
and that of the massless and massive mixed-symmetry fields in $\cH^{\rm\sst MixSym}_{j,\text{non-min}}$ \eqref{FF j} as
\ba
\Gamma^{\sst (1)\,\text{ren}}_{j,\text{non-min}|{\rm Sym}} \eq 
n_j\,\frac{48\,j^4-18\,j^2+1}{45}\,\log R\,,\\
\Gamma^{\sst (1)\,\text{ren}}_{j,\text{non-min}|\rm MixSym} +\Gamma^{\sst (1)\,\text{ren}}_{j,\text{non-min}|\rm Massive}\eq
n_j\,\frac{72\,j^4-42\,j^2+1}{45}\,\log R\,.
\ea
In principle, we can also consider the case where the boundary theory 
is given by only positive (or negative) helicity $j$, hence
$\cS_{[j,0]}$ (or $\cS_{[0,j]}$). 
Then, the unitarity contrains the boundary 
fields to be complex: it can have $U(N)$ symmetry but not $O(N)$.
If we apply the vectorial duality to this chiral model, then the field content
of the bulk (non-minimal) theory would have the spectrum
\be
    \cH_{[j,0],\text{non-min}}= \cS_{[j,0]}\otimes\cS_{[0,j]}
    =\frac12\,\cH_{j,\text{non-min}|\rm Sym}\,,
\ee
where $\cS_{[0,j]}$ is for the complex conjugate of the chiral fields 
in $\cS_{[j,0]}$\,.
Therefore, its vacuum energy will be given by
\be
\Gamma^{\sst (1)\,\text{ren}}_{[j,0],\text{non-min}}
=\frac12\,\Gamma^{\sst (1)\,\text{ren}}_{j,\text{non-min}|\rm Sym}
=n_j\,\frac{48\,j^4-18\,j^2+1}{90}\,\log R\,,
\ee
In contrast to the parity invariant model,
the above vacuum energy is not proportional to that of $\cS_{[j,0]}$,
\be
   \Gamma^{\sst (1)\,\text{ren}}_{\cS_{[j,0]}}=
   \left(-1\right)^{2j}\,\frac{60\,j^4-30\,j^2+1}{90}\,\log R\,,
\ee 
with an integer factor
except for the trivial cases of  $j=0$ and ${1\over 2}$
where the prefactor $n_j$ itself vanishes.

\subsubsection*{Minimal Type-$j$ Theory}

We now turn to the minimal type-$j$ theory, which is the putative dual of the $O(N)$ vector model built from spin-$j$ doubleton. The character of minimal type-$j$ theory is given by
\begin{equation}
\chi^{\phantom{g}}_{j,\text{min}}\!\left(\beta,\alpha_+,\alpha_-\right) = {\chi^{\phantom{g}}_{\cS_j}(\beta,\alpha_+,\alpha_-)^2+(-1)^{2j}\, \chi^{\phantom{g}}_{\cS_j}\!\left(2\,\beta,2\,\alpha_+,2\,\alpha_-\right)\over 2},
\label{min char}
\end{equation}
where $\cS_{j}$ is the parity invariant spin-$j$ singleton. The contribution to the vacuum energy from the first term 
is the half of the non-minimal theory, already computed in \eqref{1lvacjnonmin}. We will work with the second term, 
$\chi_{h}\!\left(\beta,\alpha_+,\alpha_-\right)=(-1)^{2j}\,\frac12\,
\chi^{\phantom{g}}_{\cS_j}\!\left(2\,\beta,2\,\alpha_+,2\,\alpha_-\right)$. Also, for brevity we directly write the expressions for the $\gamma_{h|n}$'s:
\begin{equation}
\begin{split}
\gamma_{h|2} &= \frac{(-1)^{2j}}{960} \left(480\, j^4-240\, j^2+13\right),\\
\gamma_{h|1} &= \frac{(-1)^{2j}}{576} \left(192\, j^4-96\, j^2+5\right),\\
\gamma_{h|0} &= (-1)^{2j}\,\frac{2\,j^4-j^2}{4}\,.
\end{split}
\end{equation}
These may in turn be summed to obtain the contribution of this term to the one-loop vacuum energy,
\begin{equation}
\Gamma^{\sst (1)\,\text{ren}}_{h} =
(-1)^{2j}\,\frac{1}{45} \left(60\,j^4-30\,j^2+1\right)\log R\,.
\end{equation}
We observe that this exactly matches with the corresponding answer for $\cS_j$\,. 
Finally, the one-loop vacuum energy of the minimal Type-$j$ model may be expressed as
\begin{equation}
\Gamma^{\sst (1)\,\text{ren}}_{j,\text{min}} ={1\over 2}\,\Gamma^{\sst (1)\,\text{ren}}_{j,\text{non-min}}+\Gamma^{\sst (1)\,\text{ren}}_{h}=
\left[\left(-1\right)^{2j}\,n_j+1\right]
\Gamma^{\sst (1)\,\text{ren}}_{\cS_j}\,.
\label{min VE}
\end{equation}
Compared to the non-minimal theory case \eqref{1lvacjnonmin}, we find  that
the relation between the vacuum energy of type-$j$ theory and that of spin-$j$ doubleton
has an additional term `+1'. The latter contribution is responsible a shift of bulk coupling constant
proposed in the lower $j$ cases.

\subsection{One-Loop Casimir Energy in AdS$_5$ with $S^1\times S^3$ boundary}
\label{sec:TAdS}

When the background is the thermal AdS$_5$ with $S^1\times S^3$ boundary,
the one-loop partition function has two contributions,
\be
	\G^{\sst (1)\,\rm ren}_{\cH}(\b)
	=\b\,\cE_{\cH}+\hat\cF_{\cH}(\b)\,,
\ee
where $\b$ is the radius of $S^1$ and  the Casimir energy $\cE_{\cH}$ is given by 
a contour integral of the character,
\be
	\cE_{\cH}
	=-(-1)^F\,\frac{1}{2}\,\oint \frac{d\b}{2\,\pi\, i\,\b^2}\,\chi_\cH(\b,0,0)\,,
	\label{E H}
\ee
where $F$ is the fermion number
for $\cH$\,.
The contour can be taken as a circle around the origin and the residue theorem can be applied \cite{Bae:2016hfy}. 
Thus, $\cE_{\cH}$ is given by $-{1\over 2}$ times the coefficient of the $\beta^1$ term in the Laurent  expansion of $\chi_{\cH}\!\left(\b\right)$ at $\beta=0$\,.

As we mentioned in Introduction, the boundary $S^1\times S^3$
is not a compatible space for massless fields with spin $j\ge 3/2$ 
as the latter can be defined only in a conformally flat and Einstein background \cite{Christensen:1978md,Aragone:1979hx},
whereas $S^1\times S^3$ is not Einstein.
Nevertheless, one can consider the field content of type-$j$ theory in the thermal AdS$_5$
and calculate its Casimir energy. 
Even though we expect the type-$j$ holography in thermal AdS to be inconsistent,
we present the calculation of its Casimir energy for completeness.
Indeed, one can see that the result shows an important difference from the $S^4$ boundary case.
This calculation has been already presented  in \cite{Gunaydin:2016amv} (while the present work has been almost completed), 
and various problematic features of the theory are discussed.
For completeness, we reproduce
the result here.

For the computation of the Casimir energy, it is again sufficient to know the (blind) character of the system,
which is nothing but the $\a_1=\a_2=0$ case of the character \eqref{Sj char} and given by 
\be
	\chi_{\cS_j}(\b,0,0)
	=e^{-j\,\b}\,\frac{2\,j\,(\cosh\b-1)+\sinh\b}{(\cosh\b-1)^2}\,.
	\label{bl char}
\ee
From the above, it is sufficient to extract its $\b$-linear Laurent coefficient
to get the residue. In this way, we get
\be\label{ecsj}
	\cE_{\cS_j}=\left(-1\right)^{2j}\,\frac{30\,j^4-20\,j^2+1}{120}\,.
\ee
In principle, this can be interpreted as the Casimir energy of a single massless spin-$j$
if we ignore the fact that the latter is not well-defined around $S^1\times S^3$\,.
We can proceed to the non-minimal type-$j$ theory,
whose character is simply the square of \eqref{bl char} as in \eqref{non min char}:
\be
	\chi_{j,\text{non-min}}(\b,0,0)
	=e^{-2j\,\b}\,\frac{\left[2\,j\,(\cosh\b-1)+\sinh\b\right]^2}{(\cosh\b-1)^4}\,,
\ee
and from its $\b$-linear coefficient, we conclude that
\be\label{ecsnm}
	\cE_{j,\text{non-min}}=n_j\,{288\,j^4-208\,j^2-3\over 420}\,.
\ee
The blind character of the minimal theory is defined as in \eqref{min char},
from which one can extract the Casimir energy as
\be
\cE_{j,\text{min}} =n_j\,{288\,j^4-208\,j^2-3\over 840}+(-1)^{2j}\,\frac{30\,j^4-20\,j^2+1}{120}\,.
\ee
We see that apart from the familiar `low spin' cases of $j=0$, $j=\frac12$ \cite{Giombi:2014yra}, and $j=1$ \cite{Beccaria:2014zma}, there is no value of $j$ for which the Casimir energy $\cE_{j,\text{non-min}}$
or  $\cE_{j,\text{min}}$ is an integer times of $\cE_{\cS_j}$\,.
This is to be contrasted to the results \eqref{1lvacjnonmin}  and \eqref{min VE} in the $S^4$ boundary case.

\subsection{Extension to Type AZ Theory}\label{sec:aomegaHS}

We consider now a more speculative model where
 the boundary theory itself is given by an infinite number of massless higher spin fields.
More precisely, we take it as the collection of free massless spin-$j$ fields with $j=0,1,2,\ldots,\infty$\,.
This spectrum coincides with that of  non-minimal Vasiliev theory 
but now it is placed on the boundary
(see \cite{Vasiliev:2001zy} for related discussions). The boundary theory carries the reducible representation,
\be
	\mathcal{S}_{{\rm AZ}}=\cS_{[0,0]}\oplus \bigoplus_{j=1}^\infty \left(\cS_{[j,0]}\oplus \cS_{[0,j]}\right),
\ee
and the corresponding character can be obtained by summing the spin-$j$ characters \eqref{Sj char} over all integers $j$ as
\ba
&&\chi^{\phantom{a}}_{\mathcal{S}_{{\rm AZ}}}(\b,\a_1,\a_2)
=
\chi^{\phantom{a}}_{\mathcal{S}_{[0,0]}}
 +\sum_{j=1}^{\infty}\left(\chi^{\phantom{a}}_{\mathcal{S}_{[j,0]}}+
 \chi^{\phantom{a}}_{\mathcal{S}_{[0,j]}}\right)\nn
 &&=\,
 \frac{ 1 + \cos\alpha_1 + \cos\alpha_2 + \cosh\beta }{1 + \cos(2\alpha_1) + \cos(2\alpha_2)+ \cosh(2\b)
 -4\,\cos\alpha_1 \cos\alpha_2\, \cosh\b}.
\ea
Here, one needs to be careful in treating the $j=0$ contribution as it is already parity invariant.
The field-theory description is nothing but
the collection of Fronsdal fields on the boundary:
\be
    S_{\rm CFT}=\int d^4x\,\sum_{i=1}^N \sum_{j=0}^\infty \bar \phi_i^{a_1\cdots a_j}\,
    G_{i\,a_1\cdots a_j}\,.
\ee
Again, we can put these fields either in $U(N)$ or $O(N)$ vector representation. In the latter case, the Fronsdal fields become real. 

As the boundary fields carry a vector representation, all single trace operators are given as bilinear gauge invariants.
Again, the content of the latter operators can be easily identified by taking tensor
products of the representation $\cR$ carried by the boundary theory.
This operator content corresponds to the field content of the bulk theory, which we shall refer to as 
type AZ theory.
When the boundary theory carries $U(N)$ vector representation, 
its single trace operator spectrum is given by
\be \label{decomposeAZ}
	\cH_{{\rm AZ},U(N)}=\mathcal{S}_{{\rm AZ}}\otimes\mathcal{S}_{{\rm AZ}}\,.
\ee
Since each $\mathcal{S}_{{\rm AZ}}$ contains all spins, its product contains
all possible products \cite{Dolan:2005wy,Beccaria:2014zma}\,:
\ba
	S_{[j,0]}\otimes S_{[j',0]}
	\eq \bigoplus_{k=|j-j'|}^{j+j'-1} \cD(j+j'+2,[k,0])\oplus
	\bigoplus_{k=0}^\infty\cD(j+j'+2+k;[j+j'+\tfrac k2,\tfrac k2])\,, \nn
	S_{[j,0]}\otimes S_{[0,j']}
	\eq \bigoplus_{k=0}^\infty \cD(j+j'+2+k;[j+\tfrac k2,j'+\tfrac k2])\,.
	\label{jj'}
\ea
Here, we again split the spectrum of $\cH_{{\rm AZ},U(N)}$ into three parts:
\be 
	\cH_{{\rm AZ},U(N)}
	=\cH^{\rm\sst Sym}_{{\rm AZ},U(N)}\oplus \cH^{\rm\sst MixSym}_{{\rm AZ},U(N)}
	\oplus \cH^{\rm\sst Massive}_{{\rm AZ},U(N)}\,.
	\label{AZ decomp}
\ee
The first part contains symmetric conserved currents,
\be
    \cH^{\rm\sst Sym}_{{\rm AZ},U(N)}=
    \bigoplus_{s=0}^\infty (2[s/2]+1)\,
	\cD(s+2,[\tfrac s2,\tfrac s2])\,,
	\label{sym AZ}
\ee
which are dual to massless symmetric fields in AdS$_5$. One can see that there are growing number of fields 
as the spin increases. The second and third parts contain the mixed-symmetry conserved currents
dual to massless mixed symmetry fields
and the long operators
dual to non-gauge two-row fields, respectively. 
The precise multiplicities of these fields
in the type AZ theory can be identified from \eqref{jj'}.
One can obtain analogously the single-trace operator content for $O(N)$ model
which is dual to minimal type AZ theory in the bulk.

We now turn to computing the one-loop vacuum energy of type AZ theory in AdS$_5$. 
Following the CIRZ method, it is sufficient to identify the underlying character of the theory.
In the case of the non-minimal theory, dual to the $U(N)$ vector model, the character is given by 
\begin{equation}
\chi^{\phantom{a}}_{\text{AZ,non-min}}(\b,\a_1,\a_2) = \left[\chi_{\mathcal{S}_{\rm AZ}}(\b,\a_1,\a_2)\right]^2\,.
\label{nm AZ}
\end{equation}
The next step is to calculate $f_{\text{AZ,non-min}|n}$ using \eqref{f H}. 
Explicit expressions encountered in the derivation are fairly lengthy, but the final form taken by these quantities is compact, and given by
\begin{equation}
\begin{split}
f_{\text{AZ,non-min}|2}\!\left(\beta\right) &= \frac{1}{128}\,(\cosh\beta+3)^2\,
 \text{csch}^4\tfrac{\beta }{2}\,,\\
f_{\text{AZ,non-min}|1}\!\left(\beta\right) &= \frac{1}{768}\, (\cosh\beta+3) (64\, \cosh\beta+\cosh (2 \beta )+79)\, 
\text{csch}^6\tfrac{\beta }{2}\,,\\
f_{\text{AZ,non-min}|0}\!\left(\beta\right) &= \frac{1}{64}\,(115\, \cosh\beta+21\,\cosh (2 \beta )+\cosh (3 \beta )+103)\, \text{csch}^8\tfrac{\beta }{2}\,.
\end{split}
\end{equation}
Then by series expanding about $\beta=0$ and extracting the appropriate Laurent coefficients, we find
that  the coefficients $\g_{\text{non-min}|n}$ \eqref{g H} all vanish:
\begin{equation}
\gamma_{\text{AZ,non-min}|2}= \gamma_{\text{AZ,non-min}|1}= 
\gamma_{\text{AZ,non-min}|0}= 0\,.
\end{equation}
Therefore, the full one-loop vacuum energy of non-minimal type AZ theory vanishes:
\begin{equation}\label{azvac}
\Gamma^{\rm\sst (1)\,\text{ren}}_{\text{AZ,non-min}} = 0\,.
\end{equation}
If we decompose
the above full vacuum energy
into the contributions from symmetric 
and massless and massive mixed-symmetry fields as (\ref{AZ decomp}),
each one-loop vacuum energy 
does not vanish:
\begin{equation}
\Gamma^{\rm\sst (1)\,\text{ren}\,}_{\text{AZ,non-min}|\text{Sym}} = 
\frac{1033}{90720} 
\,\log R\,,\, \qquad \Gamma^{\rm\sst (1)\,\text{ren}\,}_{\text{AZ,non-min}|\text{MixSym}} +\Gamma^{\rm\sst (1)\,\text{ren}\,}_{\text{AZ,non-min}|\text{Massive}}= 
-\frac{1033}{90720}\,\log R\,,
\end{equation}
but they cancel each other to give \eqref{azvac}.

One can repeat the computation for the minimal theory, dual to $O(N)$ vector model, by using the character,
\begin{equation}
\chi^{\phantom{a}}_{\text{AZ,min}}(\b,\a_1,\a_2) =
\frac{ \left[\chi_{\mathcal{S}_{\rm AZ}}(\b,\a_1,\a_2)\right]^2+\chi_{\mathcal{S}_{\rm AZ}}(2\,\b,2\,\a_1,2\,\a_2)}2\,.
\label{min AZ}
\end{equation}
After straightfoward calculations, we find again that all the $\gamma_{\text{AZ,min}|n}$
coefficients vanish, hence so does the full one-loop vacuum of minimal type AZ theory:
\begin{equation}
\gamma_{\text{AZ,min}|2}=\gamma_{\text{AZ,min}|1}= 
\gamma_{\text{AZ,min}|0}=0\,,
\qquad 
\Gamma^{\sst (1)\,\text{ren}}_{\text{AZ,min}}=0\,.
\end{equation}
In fact, there is a simple reason for that all these $\gamma_{\cH|n}$ coefficients vanish (hence the vacuum energy).
It is because the underlying character $\chi^{\phantom{a}}_{\cS_{\rm AZ}}$
is even in $\b$\,:
\be
	\chi^{\phantom{a}}_{\cS_{\rm AZ}}(-\b,\a_1,\a_2)=\chi^{\phantom{a}}_{\cS_{\rm AZ}}(\b,\a_1,\a_2)\,.
\ee
This property is preserved in making the characters for non-minimal \eqref{nm AZ} and minimal \eqref{min AZ} theories.
Moreover, if the character is invariant in $\b\leftrightarrow -\b$, then the corresponding $f_{\cH|n}(\b)$'s are even in $\b$\,. Since $\g_{\cH|n}$ correspond to Laurent coefficients of odd $\b$ powers, they simply vanish for any even function $f_{\cH|n}(\b)$,  hence as a result, the one-loop vacuum energy vanishes.

A priori, the collection of all massless integer spins cannot be better defined on $S^1\times S^3$ than
its individual spin part. Nevertheless, we can compute the Casimir energy of 
type AZ theory by placing its field content in the thermal AdS$_5$\,.
The computation of Casimier energy requires only the blind character, which has again very simple form,
\be
\chi^{\phantom{a}}_{\mathcal{S}_{\rm AZ}}\!\left(\b,0,0\right) 
= \frac{1}{8}\left(\cosh\b+3 \right) \csch^4\tfrac{\b}{2}\,.
\ee
Like the full character itself, the above blind character is even in $\b$ and this property guarantees that
Casimir energies for boundary theory, and non-minimal and minimal bulk theories are zero:
\begin{equation}
	\cE_{\mathcal{S}_{\rm AZ}} = 0\,,\quad \cE_{\text{AZ,non-min}} =0\,, \quad \cE_{\text{AZ,min}} =0\,.
\end{equation}
It is not clear though whether this result suggests that the type AZ theory has any better chance to be well-defined
in the thermal AdS$_5$\,.

\section{Discussion}
\label{conclusion}

Let us recapitulate the type-$j$ one-loop results we obtained in this paper.
The AdS$_5$ vacuum energies are
\be
\Gamma^{\sst (1)\,\text{ren}}_{j,\text{non-min}} 
=(-1)^{2j}\,n_j\,2\,\Gamma^{\sst (1)\,\text{ren}}_{\cS_j}\,,
\qquad \Gamma^{\sst (1)\,\text{ren}}_{j,\text{min}} =\left[(-1)^{2j}\,n_j+1\right]
\Gamma^{\sst (1)\,\text{ren}}_{\cS_j}\,,
\ee
where $n_j$ is an integer and $\G^{\sst (1)\,\text{ren}}_{\cS_j}$
is the vacuum energy of spin-$j$ doubleton treated as if it is a AdS$_5$ field:
\begin{equation}
n_j=\frac{\left(2\,j-1\right)2\,j\left(2\,j+1\right)}6\,,
\qquad
\Gamma^{\sst (1)\,\text{ren}}_{\cS_j} =(-1)^{2j}\, \frac{60\,j^4-30\,j^2+1}{45}\,\log R\,.
\label{n G}
\end{equation}
Let us first emphasize that it is a non-trivial fact that the one-loop result of the bulk theory 
is related to that of doubleton in this special fashion:
this does not happen neither in (anti-)chiral model case nor in the Casimir energy computations.
In the type A, B, C models with $j=0, \frac12, 1$\,,
the  vacuum energy $\G^{\sst (1)\,\text{ren}}_{\cS_j}$\,,
\be
	\left(\G^{\sst (1)\,\text{ren}}_{\cS_0}\,,\,\G^{\sst (1)\,\text{ren}}_{\cS_{\frac12}}
	\,,\,\G^{\sst (1)\,\text{ren}}_{\cS_1}\right)
	=\left(\frac1{90}\,,\,\frac{11}{180}\,,\,\frac{31}{45}\right) \log R\,,
\ee
happens to be related to the free energy of the boundary theory on $S^4$ (with $N=1$),
\be
	\left(F_0\,,\,F_{\frac12}\,,\,F_1\right)
	=\left(\frac1{90}\,,\,\frac{11}{180}\,,\,\frac{31}{45}\right) \log \L_{\sst\rm CFT}\,,
\ee
which are proportional to the $a$-anomaly coefficients.
This result, upon the IR/UV identification, $\log R=\log \L_{\sst\rm CFT}$\,,
suggests that the inverse coupling constant of the bulk theory be related to $N$ with a certain integer shift:
\be
	g^{-1}_{\text{non-min}}=N-(-1)^{2j}\,n_{j}\,,\qquad g^{-1}_{\text{min}}=N-(-1)^{2j}\,n_{j}-1\qquad
	[j=0,\tfrac12,1]\,,
	\label{j shift}
\ee
with $n_j$ as defined in \eqref{n G}.
Now the question is whether the vacuum energy $\G^{\sst (1)\,\text{ren}}_{\cS_j}$
can be analogously related to the $S^4$ free energy $F_j$ of massless spin-$j$ for higher values of $j$\,. If so, we would expect the the dictionary \eqref{j shift} to continue to hold for these theories as well.
In the following, we examine this possibility.

Let us now turn to the 
free energy of massless integer spin $j$ field over $S^4$ of unit radius. It is given by
\begin{equation}\label{zs4}
       F_j =\frac12\left[
       \log\det\left(-\Box-\left(j^2-2j-2\right)\right)_{(j)}-
       \log\det\left(-\Box-\left(j^2-1\right)\right)_{(j-1)}\right],
\end{equation}
where $-\Box$ is the Laplace operator on $S^4$ with positive definite eigenvalues and the subscript $(j)$ denotes that the operator acts on a rank $j$ symmetric transverse traceless (STT) tensor. The multiplicities $d_n(j)$ and the eigenvalues of the kinetic operators are given by $\l_n(\D,j)$\,\cite{Gibbons:1978ji}:
\ba
    && d_n(j)={\rm dim}(n+j,j)=
    \frac16\,(2j+1)\left(n+1\right)\left(2j+2+n\right)\left(2j+3+2n\right),\nn
    &&
    \lambda_n(\D,j)=(n+j+\D)(n+j+3-\D)\,,\quad n\geq 0.
    \label{d lambda}
\ea
Focussing on the logarithmically divergent part, the $S^4$ free energy is given by
\be
	F_{j}=-\left(\zeta_{j+1,j}(0)-\zeta_{j+2,j-1}(0)-\eta_{j}\right) \log\L_{\sst\rm CFT}\,,
	\label{F j}
\ee
where $\eta_j$ is the contribution due to zero modes and the zeta function $\zeta_{\D,j}$ for
an irreducible field labeled by $(\D,j)$, is given by
\be
     \zeta_{\D,j}(z)=\sum_{n=0} d_n(j)\,\lambda_n(\D,j)^{-z}\,,
\ee
Here, we are interested only in the logarithmically divergent part of $F_{j}$\,,
so we need to idenify $\zeta_{j+1,j}(0)-\zeta_{j+2,j-1}(0)$ and $\eta_j$\,.
The former can be readily obtained  as
\be\label{zeta0nz}
   \zeta_{j+1,j}(0)-\zeta_{j+2,j-1}(0)
    =\frac{15\,j^2-1}{45}\,.
\ee
Therefore, if the zero-mode contribution $\eta_j$ is the desired quantity $\eta_j^{\sst\rm desired}$\,:
\be
	\eta_j^{\sst\rm desired}=\frac{j^2\,(4\,j^2-1)}{3}=j\,n_j\,,
	\label{zero j}
\ee
then the free energy $F_j$ would coincide with $\G^{\sst (1)\,\text{ren}}_{\cS_j}$,
and the dictionary \eqref{j shift} would make a sense for any $j$\,.

 In section 2 of  \cite{Tseytlin:2013fca}, Tseytlin analyzed the 
  zero modes for massless spin-$j$\,: 
  they arise when we decompose a
 traceless rank-$j$ tensor $\phi^{\sst\rm T}_{(j)}$ into 
 the traceless and transverse part $\phi^{\sst\rm TT}_{(j)}$ and the rest:
 \be
 	\phi^{\sst\rm T}_{\mu_1\cdots \mu_j}
 	=\phi^{\sst\rm TT}_{\mu_1\cdots\mu_j}
 	+\P\left[\nabla_{(\mu_1}\,\xi^{\sst\rm T}_{\mu_2\cdots \mu_j)}\right].
\ee
The rest is the traceless part ($\Pi$ is
the traceless projector) of gradient of a traceless tensor $\xi^{\sst\rm T}_{(j-1)}$\,. In order that this decomposition is one-to-one,
$\xi^{\sst\rm T}_{(j-1)}$ should not involve the solutions of
$\P[\,\nabla\,\xi^{\sst\rm T}_{(j-1)}\,]=0$\,, namely zero modes.
These precisely correspond to the spin-$j$ conformal Killing tensors, whose number is 
$j^2(j+1)^2(2j+1)/12$\,. In the analysis of massless spin-$j$, we face this decomposition twice,
once for the rank-$j$ physical mode and the other time for the rank-$(j-1)$ gauge mode. Hence,
according to \cite{Tseytlin:2013fca}, the total zero-mode contribution is
\be
	\eta^{\rm\sst Tseytlin}_j=\frac{j^2\,(j+1)^2\,(2j+1)}{12}-\frac{(j-1)^2\,j^2\,(2j-1)}{12}
	=\frac{5\,j^4+j^2}6\,,
	\label{T zm}
\ee
and consequently, the free energy with the zero-mode contribution \eqref{T zm},
\be
	F_j=\frac{75 j^4-15 j^2+2}{90}\,\log\L_{\rm\sst CFT}\,,
	\label{T F}
\ee
differs from the vacuum energy $\G^{\sst (1)\,\rm ren}_{\cS_j}$ \eqref{n G}.
Let us make one curious observation:
if we neglect the $\gamma_{\cH|0}$ term \eqref{gamma S j} --- which is non-vanishing only for two-row Young diagram tensors --- in the $\G^{\sst (1)\,\rm ren}_{\cS_j}$ computation, we would get the result \eqref{T F}.

Differently from the renormalized quantity, the logarithmically divergent part of free energy may depend on field-theoretical realizations \cite{Duff:1980qv}. Therefore, in principle, there might be other formulation
of massless spin-$j$  than the Fronsdal one with the free energy given by $\G^{\sst (1)\,\rm ren}_{\cS_j}$\,.
For instance, if one considers
the Maxwell-like formulation \cite{Skvortsov:2007kz,Campoleoni:2012th,Francia:2013sca} where
the gauge field is traceless and the gauge parameter is traceless and traceverse, we need to consider
the decomposition,
 \be
 	\phi^{\sst\rm T}_{\mu_1\cdots \mu_j}
 	=\phi^{\sst\rm TT}_{\mu_1\cdots\mu_j}
 	+\nabla_{(\mu_1}\,\xi^{\sst\rm TT}_{\mu_2\cdots \mu_j)}\,,
\ee
only once, hence the zero modes only appear here.
They are the solutions of $\nabla\,\xi^{\sst\rm TT}_{(j-1)}=0$\,,
namely spin-$j$ Killing tensors.
These also coincide with the zero modes 
of the gauge sector $(\D=j+2,s=j-1)$ in \eqref{d lambda} and they correspond to  $n=0$ modes with the multiplicity,
\be\label{nzeroapp}
	d_0(j-1)=\frac{j\,(4\,j^2-1)}3=n_j\,.
\ee
Still this number is not sufficient to give the desired contribution
\eqref{zero j}, but misses a factor of $j$\,.
Hopefully, there may be yet another formulation of massless spin-$j$ field
which gives the desired zero modes.

As a final note, we also point out that 
the above zero mode analysis
may be affected by 
their non-trivial scalings,
which have been pointed out
in \cite{Banerjee:2010qc,Banerjee:2011jp,Sen:2011ba,Sen:2012cj,Bhattacharyya:2012wz,Bhattacharyya:2012ye}.
In the Appendix, we discuss how the inclusion of such scalings may alter possibly the results of one-loop free energy computations on $S^4$ so as to match the result with $\Gamma^{\sst (1)\,\rm ren}_{\mathcal{S}_j}$.

\acknowledgments

We are grateful to 
Prarit Agarwal,
Eduardo Conde, 
Rajesh Gopakumar,
Imtak Jeon,
Karapet Mkrtchyan,
Prithvi Narayan
and 
Ashoke Sen
for useful discussions. 
The work of EJ was supported in part by the National Research Foundation of Korea through the grant NRF2014R1A6A3A04056670 and the Russian Science Foundation grant 14-42-00047 associated with Lebedev Institute. 
The work of SL is supported by the Marie Sklodowska Curie Individual Fellowship 2014. SL thanks the Korea Institute for Advanced Study for hospitality while part of this work was carried out.

\section*{Appendix}
\appendix
\section{Free Energy of Fronsdal Fields on $S^4$}
We begin with a discussion of a generality that has been reviewed many times in the context of computing quantum corrections in AdS spaces  \cite{Banerjee:2010qc,Banerjee:2011jp,Sen:2011ba,Sen:2012cj,Bhattacharyya:2012wz,Bhattacharyya:2012ye,Gupta:2013sva,Gupta:2014hxa}.  
Given the one-loop determinant of a theory, for definiteness taken to be in four dimensions, we have
\begin{equation}
\mathcal{Z}_{1-\ell} = \int \left[\mathcal{D}\phi\right] e^{-\int d^4x \,\phi\,\mathcal{O}\,\phi} =  \left(\det\!'\mathcal{O}\right)^{-1/2}\times\mathcal{Z}_{zero},
\end{equation}
where by $\det\!'$ we mean the determinant evaluated over the non-zero modes of $\mathcal{O}$ and $\mathcal{Z}_{zero}$ is the path integral over the zero modes of $\mathcal{O}$. 
Next, we assume that the theory is defined on a manifold with a length scale $a$ and we are computing the part of the free energy $\ln\mathcal{Z}_{1-\text{loop}}$ proportional to $\ln a$. 
It was observed in \cite{Banerjee:2010qc,Banerjee:2011jp} that the zero mode associated with a given field $\phi$ in the spectrum of the theory may scale with $a$ differently compared to the non zero modes. In particular, suppose that 
the contribution of a single zero mode of the field $\phi$ scales as 
\begin{equation}\label{betadef}
\mathcal{Z}_{zero}\sim a^{\beta_{\phi}}\mathcal{Z}_{o},
\end{equation}
where $\mathcal{Z}_o$ does not scale with $a$. Then if the field $\phi$ has $n_\phi$ number of zero modes then
\begin{equation}
F=-\ln\mathcal{Z}_{1-\ell} = -\left(\tilde{\zeta}\left(0\right) + n_\phi\,\beta_\phi\right)\ln a +\ldots,
\end{equation}
where $\ldots$ denote terms that are not proportional to $\ln a$, and $\tilde{\zeta}$ is the zeta function evaluated over the non-zero modes of the field $\phi$. In practice, it is often easier to compute the zeta function over all possible modes, denoted by $\zeta$ including zero modes, and use the equivalent expression\footnote{In particular, $\tilde{\zeta}\left(0\right) = \zeta\left(0\right)-n_\phi$.} \cite{Banerjee:2010qc,Banerjee:2011jp}
\begin{equation}
F=-\ln\mathcal{Z}_{1-\ell} = -\left(\zeta\left(0\right) + n_\phi\left(\beta_\phi-1\right)\right)\ln a +\ldots.
\end{equation}
Note that generically, $\beta_\phi$ is not equal to one. We refer the reader to \cite{Banerjee:2011jp,Sen:2012cj,Bhattacharyya:2012ye} for explicit examples in various dimensions.

The eigenvalues and degeneracies of the kinetic operators in \eqref{zs4} have been enumerated in \eqref{d lambda}. Using those expressions we see that for the expression \eqref{zs4} it is the ghost determinant acting over spin $j-1$ STT tensors which has a zero eigenvalue for the lowest quantum number $n=0$. The number of such zero modes is given \eqref{nzeroapp}. Physically, since the zero modes belong to the ghost path integral, they should be interpreted as the part of the gauge symmetry $\delta\phi_{(j)} = \nabla\,\xi_{(j-1)}$ that remains even after fixing gauge \cite{Volkov:2000ih,Bhattacharyya:2012ye}. Next, we turn to the computation of the free energy from the one-loop determinants \eqref{zs4} and get \eqref{F j} with
\be
    \eta_j=n_j\,\left(\beta_{j}-1\right),
    \label{eta j b}
\ee
where 
$n_{j}$ is exhibited in \eqref{nzeroapp}. We now turn to the computation of $\beta_{j}$ following the analysis of \cite{Banerjee:2010qc,Banerjee:2011jp,Sen:2011ba,Sen:2012cj,Bhattacharyya:2012wz,Bhattacharyya:2012ye, Gupta:2015gga} who worked in Anti-de Sitter space. 
We start with the normalization for the path integral for the ghost determinant for a spin-$j$ Fronsdal field on $S^4$, i.e.
\begin{equation}
\int \left[\mathcal{D}\xi_{\mu_2\ldots\mu_j}\right]\exp{\left[-\int d^{4}x\,\sqrt{g}\, g^{\mu_2\nu_2}\,\ldots\, g^{\mu_j\nu_j}\, \xi_{\mu_2\ldots\mu_j}\,\xi_{\nu_2\ldots\nu_j}\right]}=1.
\end{equation}
Next, if the radius of the $S^4$ is $a$ then the metric $g$ scales as $g_{\mu\nu} = a^2\,g^{(0)}_{\mu\nu}$. As a result, the normalization becomes
\begin{equation}
\int \left[\mathcal{D}\xi_{\mu_2\ldots\mu_j}\right]\exp{\left[-a^{4-2(j-1)}\int d^{4}x\,\sqrt{g^{(0)}}\, g^{(0)\mu_2\nu_2}\,\ldots\, g^{(0)\mu_j\nu_j}\, \xi_{\mu_2\ldots\mu_j}\,\xi_{\nu_2\ldots\nu_j}\right]}=1.
\end{equation}
Hence the correctly normalized integration measure is given by
\begin{equation}
\left[\mathcal{D}\xi_{\mu_2\ldots\mu_j}\right]=\prod_{x,(\mu_1\ldots\mu_j)}d\left(a^{3-j}\xi_{\mu_2\ldots\mu_j}\right).
\end{equation}
As we commented above, the zero modes of the ghost operator are associated with gauge transformations of the Fronsdal field $\delta\phi_{\mu_1\ldots\mu_j} = \nabla_{(\mu_1}\xi_{\mu_2\ldots\mu_j)}$, and in particular are some specific rank $j-1$ STT tensors. 
It now remains to determine the scaling of the fields $\xi_{\m_2\ldots\mu_j}$. The procedure is to determine the tensorial properties of the field $\xi$ which generate an $a$-independent symmetry algebra, and take those tensors to scale as $a^0$. 
Then, by using the metric $g_{\mu\nu}$ we may infer the scaling of $\xi_{\mu_2\ldots\mu_j}$. For this we use the fact that these are linearized transformations of a non-linear symmetry algebra which contains the isometry of $S^4$ as a subalgebra, and that the $a$-independent $S^4$-isometry is generated by the $\xi^{\mu}$. Hence, for the higher-spin algebra to be independent of $a$, we must generate it by $\xi^{\mu_2\ldots\mu_j}$, which we therefore take to scale as $a^0$. Then it follows that $\xi_{\mu_2\ldots\mu_j}$ scales as $a^{2(j-1)}$. Therefore, the scaling of each zero mode is given by $a^{2(j-1)}\times a^{3-j} = a^{j+1}$. That is, over a single zero mode
\begin{equation}
\mathcal{Z}_{zero} \sim a^{j+1}\mathcal{Z}_o.
\end{equation}
We therefore see, on comparing with \eqref{betadef} that 
\begin{equation}\label{beta}
\beta_{j} = j+1.
\end{equation}
Plugging the above in \eqref{eta j b}, we 
get the desired zero mode contribution \eqref{zero j}.

However we add a word of caution, though hyperboloids and spheres share many common features there are also potential subtle differences for example in the zero mode spectrum. For these reasons, the computations in this Appendix might be regarded as encouraging, but still preliminary and are currently under further exploration in a related context \cite{hsds4}. 

\bibliographystyle{JHEP}
\bibliography{matrix}
\end{document}